\documentclass[12pt]{article}
\usepackage{psfig}
\begin{document}
\bibliographystyle{nature}
\renewcommand{\baselinestretch}{1.66}
\normalsize
\begin{center}
\begin{Large}
{Discovery of a Supernova Explosion at Half the Age of the Universe\\
and its Cosmological Implications}

\end{Large}

\vspace{7.5pt}

S.~Perlmutter$^{1,2}$, G.~Aldering$^{1}$, M. Della Valle$^{3}$, 
S.~Deustua$^{1,4}$, R.~S.~Ellis$^{5}$,\\
S.~Fabbro$^{1,6,7}$, A.~Fruchter$^{8}$, G.~Goldhaber$^{1,2}$,
A.~Goobar$^{1,9}$, D.~E.~Groom$^{1}$,\\
I.~M.~Hook$^{1,10}$, A.~G.~Kim$^{1,11}$, M.~Y.~Kim$^{1}$,
R.A.~Knop$^{1}$, C.~Lidman$^{12}$,\\
R.~G.~McMahon$^{5}$, P.~Nugent$^{1}$, R.~Pain$^{1,6}$,
N.~Panagia$^{13}$,\\
C.~R.~Pennypacker$^{1,4}$, P.~Ruiz-Lapuente$^{14}$,
B.~Schaefer$^{15}$ \& N.~Walton$^{16}$\\
(The Supernova Cosmology Project)
\end{center}

\footnotetext[1]{E.~O. Lawrence Berkeley National Laboratory, 
Berkeley, California, USA}
\footnotetext[2]{Center for Particle Astrophysics, U.C. Berkeley,
California, USA}
\footnotetext[3]{Dipartmento di Astronomia, University de Padova, Italy}
\footnotetext[4]{Space Sciences Laboratory, U.C. Berkeley, California,
USA} 
\footnotetext[5]{Institute of Astronomy, Cambridge, United Kingdom}
\footnotetext[6]{CNRS-IN2P3, University of Paris, Paris, France}
\footnotetext[7]{Observatoire de Strasbourg, Strasbourg, France}
\footnotetext[8]{Space Telescope Science Institute, Baltimore, Maryland,
USA}
\footnotetext[9]{Department of Physics, University of Stockholm,
Stockholm, Sweeden}
\footnotetext[10]{European Southern Observatory, Munich, Germany}
\footnotetext[11]{L.P.C.C. Coll\`ege de France, Paris, France}
\footnotetext[12]{European Southern Observatory, La Silla, Chile}
\footnotetext[13]{Space Telescope Science Institute, Baltimore, Maryland, USA;
affiliated with the Astrophysics Division, 
Space Science Department of ESA}  
\footnotetext[14]{Department of Astronomy, University of Barcelona,
Barcelona, Spain}
\footnotetext[15]{Department of Astronomy, Yale University, New Haven,
Connecticut, USA} 
\footnotetext[16]{Isaac Newton Group, La Palma, Spain}

\clearpage

{\bf The ultimate fate of the universe, infinite expansion or a big
crunch, can be determined by measuring the redshifts, apparent
brightnesses, and intrinsic luminosities of very distant supernovae.
Recent developments have provided tools that make such a program
practicable: (1) Studies of relatively nearby Type Ia supernovae (SNe
Ia) have shown that their intrinsic luminosities can be accurately
determined$^{1,2,3}$; (2) New research techniques$^4$ have made it
possible to schedule the discovery and follow-up observations of
distant supernovae, producing well over 50 very distant ($z$ = 0.3 --
0.7) SNe Ia to date$^{5,6,7}$.  These distant supernovae provide a
record of changes in the expansion rate over the past several billion
years.  By making precise measurements of supernovae at still greater
distances, and thus extending this expansion history back far enough
in time, we can even distinguish$^{8}$ the slowing caused by the
gravitational attraction of the universe's mass density $\Omega_{\rm
M}$ from the effect of a possibly inflationary pressure caused by a
cosmological constant $\Lambda$.  We report here the first such
measurements, with our discovery of a Type Ia supernova (SN 1997ap) at
$z$ = 0.83.  Measurements at the Keck II 10-m telescope make this the
most distant spectroscopically confirmed supernova.  Over two months
of photometry of SN 1997ap with the Hubble Space Telescope and
ground-based telescopes, when combined with previous
measurements$^{2,5}$ of nearer SNe Ia, suggests that we may live in a
low mass-density universe. Further supernovae at comparable distances
are currently scheduled for ground and space-based observations.}

SN 1997ap was discovered by the Supernova Cosmology Project on 5 March
1997 UT, during a two-night search at the CTIO 4-m telescope
that yielded 16 new supernovae.  The search technique finds such sets
of high-redshift supernovae on the rising part of their light curves
and guarantees the date of discovery, thus allowing follow-up
photometry and spectroscopy of the transient supernovae to be
scheduled$^{4}$\nocite{pe:aigua}.  The supernova light curves were
followed with scheduled $R$-, $I$-, and some $B$-band photometry at
the CTIO, WIYN, ESO 3.6m, and INT telescopes, and with spectroscopy at
the ESO 3.6 m and Keck II telescopes.  In addition, SN 1997ap was
followed with scheduled photometry on the Hubble Space Telescope
(HST).

Figure 1 shows the spectrum of SN 1997ap, obtained on 14 March 1997 UT
with a 1.5 hour integration on the Keck II 10-m telescope.  There is
negligible ($\le$5\%) host-galaxy light contaminating the supernova
spectrum, as measured from the ground- and space-based images.  When
fit to a time series of well-measured nearby Type Ia supernova (SN~Ia)
spectra$^{9}$\nocite{sntime}, the spectrum of SN 1997ap is most
consistent with a ``normal'' SN~Ia at $z=0.83$ observed $2 \pm 2$
SN-restframe days ($\sim$4 observer's days) before the supernova's
maximum light in the restframe $B$ band.  It is a poor match to the
``abnormal'' SNe~Ia, such as the brighter SN 1991T or the fainter SN
1986G.  For comparison, the spectra of low-redshift, ``normal'' SNe~Ia
are shown with wavelengths redshifted as they would appear at $z =
0.83$.  These spectra show the time evolution from seven days before
to two days after maximum light.

Figure 2 shows the photometry data for SN 1997ap, with significantly
smaller error bars for the HST observations (Figure 2a) than for the
ground-based observations (Figure 2b and 2c).  The width of a SN~Ia's
light curve has been shown to be an excellent indicator of its
intrinsic luminosity, both at low redshift$^{1,2,3}$
\nocite{ph:delta,ha:hubble,re:lcs} and at high
redshift$^5$\nocite{7sne}: the broader and slower the light curve,
the brighter the supernova is at maximum.  We characterize this width
by fitting the photometry data to a ``normal'' SN~Ia template light
curve that has its time axis stretched or compressed by a linear
factor, called the ``stretch factor''$^{4,5}$; a ``normal'' supernova such as
SN 1989B, SN 1993O, or SN 1981B in Figure 1 thus has a stretch factor
of $s \approx 1$.  To fit the photometry data for SN 1997ap, we use
template $U$- and $B$-band light curves that have first been $1+z$
time-dilated and wavelength-shifted (``$K$-corrected'') to the $R$-
and $I$-bands as they would appear at $z=0.83$ (see ref 5 and Nugent
et~al in preparation).\nocite{7sne} The best-fit stretch factor for
all the photometry of Figure 2 indicates that SN 1997ap is a
``normal'' SN~Ia: $s=1.03\pm 0.05$ when fit for a date of maximum at
16.3 March 1997 UT (the error-weighted average of the best-fit dates
from the light curve, $15.3 \pm 1.6$ March 1997 UT, and from the
spectrum, $18 \pm 3$ March 1997 UT).

It is interesting to note that we could alternatively fit the $1+z$
time dilation of the event, holding the stretch factor constant at $s
= 1.0^{+0.05}_{-0.14}$, the best fit value from the spectral features
obtained in ref 10\nocite{nugseq95}. We find that the event lasted
$1+z = 1.86^{+0.31}_{-0.09}$ times longer than a nearby $s=1$
supernova, providing the strongest confirmation yet of the
cosmological nature of redshift$^{11,12,9}$.

The best-fit peak magnitudes for SN 1997ap are $I = 23.20 \pm 0.07$
and $R = 24.10 \pm 0.09$.  (In this letter, all magnitudes quoted or
plotted are transformed to the standard
Cousins$^{13}$\nocite{be:ubvri} $R$ and $I$ bands.)  These peak
magnitudes are relatively insensitive to the details of the fit: if
the date of maximum is left unconstrained or set to the date indicated
by the best-match spectrum, or if the ground- and space-based data are
fit alone, the peak magnitudes still agree well within errors.

The ground-based data show no evidence of host-galaxy light, but the
higher-resolution HST imaging shows a marginal detection (after
co-adding all four dates of observation) of a possible $m_I = 25.2 \pm
0.3$ host galaxy 1 arcsecond from the supernova.  This light does not
contaminate the supernova photometry from the HST and it contributes
negligibly to the ground-based photometry.  The projected separation
is $\sim$6 kpc (for $\Omega_{\rm M} = 1, \Omega_\Lambda = 0$, and $h_0
= 65$) and the corresponding $B$-band rest-frame magnitude is $M_B
\sim -17$ and its surface brightness is $\mu_B \sim 21$ mag
arcsec$^{-2}$, consistent with properties of local spiral galaxies.
We note that the analysis will need a final measurement of any
host-galaxy light after the supernova has faded, in the unlikely event
that there is a very small knot of host-galaxy light directly under
the HST image of SN 1997ap.

We compare the $K$-corrected $R$$-$$I$ observed difference of peak
magnitudes (measured at the peak of each band, not the same day) to
the $U$$-$$B$ color found for ``normal'' low-redshift SNe~Ia.  We find
that the $(U$$-$$B)_{\rm SN 1997ap} = -0.28 \pm 0.11$ restframe color
of SN 1997ap is consistent with an unreddened ``normal'' SN~Ia color,
$(U$$-$$B)_{\rm normal} = -0.32 \pm 0.12$ (see ref 14 and also Nugent
et~al. in preparation\nocite{bran:aigua}).  In this region of the sky,
there is also no evidence for Galactic reddening$^{15}$\nocite{b&h}.
Given the considerable projected distance from the putative host
galaxy, the supernova color, and the lack of galaxy contamination in
the supernova spectrum, we proceed with an analysis under the
hypothesis that the supernova suffers negligible host-galaxy
extinction, but with the following caveat:

Although correcting for $E(U$-$B) \approx 0.04$ of reddening would
shift the magnitude by only one standard deviation, $A_B = 4.8 E(U-B)
= 0.19 \pm 0.78$, the uncertainty in this correction would then be the
most significant source of uncertainty for this one supernova, due to
the large uncertainty in the $(U$$-$$B)_{\rm SN 1997ap}$ measurement
and to the sparse low-redshift $U$-band reference data.  HST
\hbox{$J$-band} observations are currently planned for future $z>0.8$
supernovae, to allow a comparison with the restframe $B-V$ color, a
much better indicator of reddening for SNe~Ia.  Such data will thus
provide an important improvement in extinction correction
uncertainties for future SNe and eliminate the need for assumptions
regarding host-galaxy extinction.  In the following analysis, we also
do not correct the lower-redshift supernovae for possible host-galaxy
extinction, so any similar distribution of extinction would partly
compensate for this possible bias in the cosmological measurements.

The significance of SNe~Ia at $z=0.83$ for measurements of the
cosmological parameters is illustrated on the Hubble diagram of Figure
3. To compare with low-redshift magnitudes, we plot SN 1997ap at an
effective restframe $B$-band magnitude of $B = 24.50 \pm 0.15$,
derived, as in ref 5,\nocite{7sne} by adding a $K$-correction and
increasing the error bar by the uncertainty due to the (small)
width-luminosity correction and by the intrinsic dispersion remaining
after this correction. By studying SNe~Ia at twice the redshift of our
first previous sample at $z \sim 0.4$, we can look for a
correspondingly larger magnitude difference between the cosmologies
considered.  At the redshift of SN 1997ap, a flat $\Omega_{\rm M} =1 $
universe is separated from a flat $\Omega_{\rm M} =0.1 $ universe by
almost one magnitude, as opposed to half a magnitude at $z \sim 0.4$.
For comparison, the uncertainty in the peak magnitude of SN1997ap is
only 0.15 mag, while the intrinsic dispersion amongst
stretch-calibrated SNe~Ia is $\sim 0.17$ mag$^5$. Thus, at such
redshifts even individual SNe~Ia become powerful tools for
discriminating amongst various world models, provided observations are
obtained, such as those presented here, where the photometric errors
are below the intrinsic SNe~Ia dispersion.

By combining such data spanning a large range of redshift, it is also
possible to distinguish between the effects of mass density
$\Omega_{\rm M}$ and cosmological constant $\Lambda$ on the Hubble
diagram$^{8}$\nocite{go:lambda}.  The blue contours of Figure 4 show
the allowed confidence region on the $\Omega_{\Lambda}$ ($\equiv
\Lambda/ (3 H_0^2$)) versus $\Omega_{\rm M}$ plane for the $z \sim
0.4$ supernovae$^5$\nocite{7sne}.  The yellow contours show the
confidence region from SN 1997ap by itself, demonstrating the change
in slope of the confidence region at higher redshift.  The red
contours show the result of the combined fit, which yields a closed
confidence region in the $\Omega_{\rm M}$-$\Omega_{\Lambda}$ plane.
This fit corresponds to a value of $\Omega_{\rm M} = 0.6 \pm 0.2 $ if
we constrain the result to a flat universe ($\Omega_{\Lambda}$ +
$\Omega_{\rm M} =1$), or $\Omega_{\rm M} = 0.2 \pm 0.4$ if we
constrain the result to a $\Lambda=0$ universe.  These results are
preliminary evidence for a relatively low-mass-density universe. The
addition of SN 1997ap to the previous sample of lower redshift
supernovae decreases the best fit $\Omega_{\rm M}$ by approximately
one standard deviation compared to the earlier results$^5$.

Our data for SN 1997ap demonstrate (1) that SNe~Ia at $z>0.8$ exist,
(2) that they can be compared spectroscopically with nearby supernovae
to determine SNe ages and luminosities and check for indications of
supernova evolution, and (3) that calibrated peak magnitudes with
precision better than the intrinsic dispersion of SNe~Ia can be
obtained at these high redshifts. The width of the confidence regions
in Figure 4 and the size of the corresponding projected measurement
uncertainties show that with additional SNe~Ia having data of quality
comparable to SN1997ap a simultaneous measurement of
$\Omega_{\Lambda}$ and $\Omega_{\rm M}$ is possible. It is important
to note that this measurement is based on only one supernova at the
highest ($z>0.8$) redshifts, and that a larger sample size is required
to find a statistical peak and identify any ``outliers.''  In
particular, SN 1997ap was discovered near the search detection
threshold and thus may be drawn from the brighter tail of a
distribution (``Malmquist bias'').  There is similar potential bias in
the lower-redshift supernovae of the Cal\'an/Tololo survey, making it
unclear which direction such a bias would change $\Omega_{\rm M}$.

Several more supernovae at comparably high redshift have already been
discovered by the Supernova Cosmology Project, including SN 1996cl,
also at $z=0.83$.  SN 1996cl can be identified as a very probable
SN~Ia, since a serendipitous HST observation (Donahue et~al., private
communication) shows its host galaxy to be an elliptical or S0. Its
magnitude and color, although much more poorly constrained by
photometry data, agree within uncertainty with those of SN 1997ap.
The next most distant spectroscopically confirmed SNe~Ia are at $z =
0.75$ and $z = 0.73$$^{16}$\nocite{pe:fourteensn97} (these supernovae
are awaiting final calibration data).  In the redshift range $z = 0.3
$ -- 0.7, we have discovered over 30 additional spectroscopically
confirmed SNe Ia, and followed them with two-filter photometry. (The
first sample of $z \sim 0.4$ SNe were not all spectroscopically
confirmed and observed with two-filter photometry$^5$.)  These new
supernovae will improve both the statistical and systematic
uncertainties in our measurement of $\Omega_{\rm M}$ and
$\Omega_{\Lambda}$ in combination.  A matching sample of $\ge$6 SNe Ia
at $z > 0.7$ is to be observed in two filters with upcoming Hubble
Space Telescope observations.  SN1997ap demonstrates the efficacy of
these complementary higher redshift measurements in separating the
contribution of $\Omega_{\rm M}$ and $\Omega_{\Lambda}$.

{\bf Acknowledgements}: We would like to thank CTIO, Keck, HST, WYIN,
ESO and the ORM -- La Palma observatories for a generous allocation of time and
the support of dedicated staff in pursuit of this project.  Dianne
Harmer, Paul Smith and Daryl Willmarth were extraordinarily helpful as
WIYN queue observers.  We would especially like to thank Gary
Bernstein and Tony Tyson for developing and supporting the Big
Throughput Camera which was instrumental in the discovery of this
supernova.

Correspondence and requests for materials should be addressed to Saul
Perlmutter (e-mail: saul@lbl.gov).

\clearpage

\renewcommand{\baselinestretch}{1.0}
\normalsize

\pagestyle{empty}
\pagenumbering{none}
\begin{figure}[tbp]
\centerline{\psfig{figure=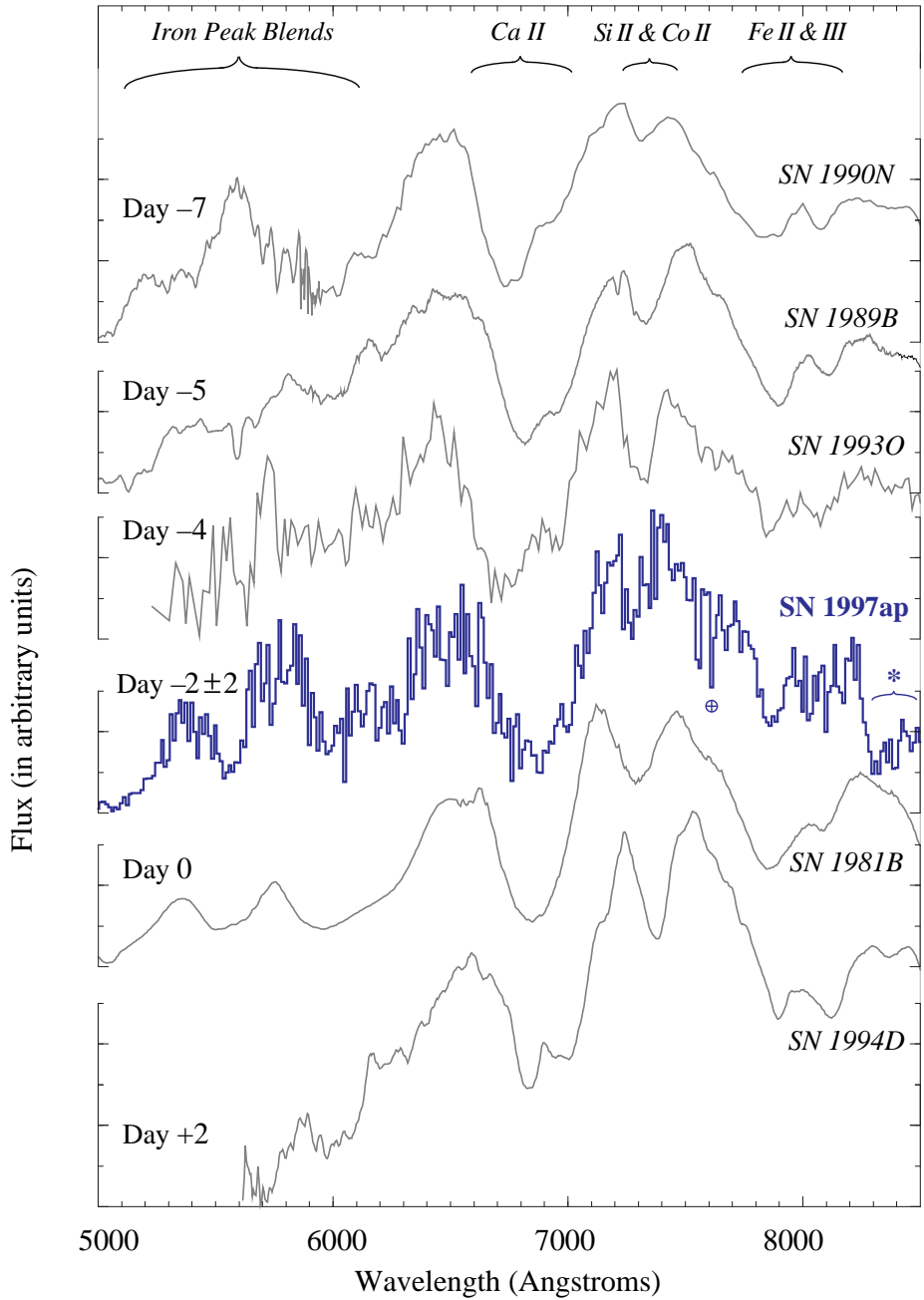,height=6.8in}}
\caption[lvsm] { Spectrum of SN 1997ap, after binning by 12.5 \AA,
placed within a time series of spectra of ``normal'' SNe~Ia
$^{17,18,19,20,21}$ (the spectrum of SN 1993O was provided courtesy of
the Cal\'an/Tololo Supernova Survey),
\nocite{le:sn1990n,we:89b,br:81b,pa:94d,iue:sn} as they would appear
redshifted to $z = 0.83$.  The spectra show the evolution of spectral
features between 7 restframe days before and 2 days after restframe
$B$-band maximum light.  SN 1997ap matches best at $2\pm 2$ days
before maximum light.  The symbol $\oplus$ indicates an atmospheric
absorption line and * indicates a region affected by night sky line
subtraction residuals. The redshift of $z = 0.83 \pm 0.005$ was
determined from the supernova spectrum itself, since there are no host
galaxy lines detected. }

\label{spectra}
\end{figure}

\begin{figure}[tbp]
\centerline{\psfig{figure=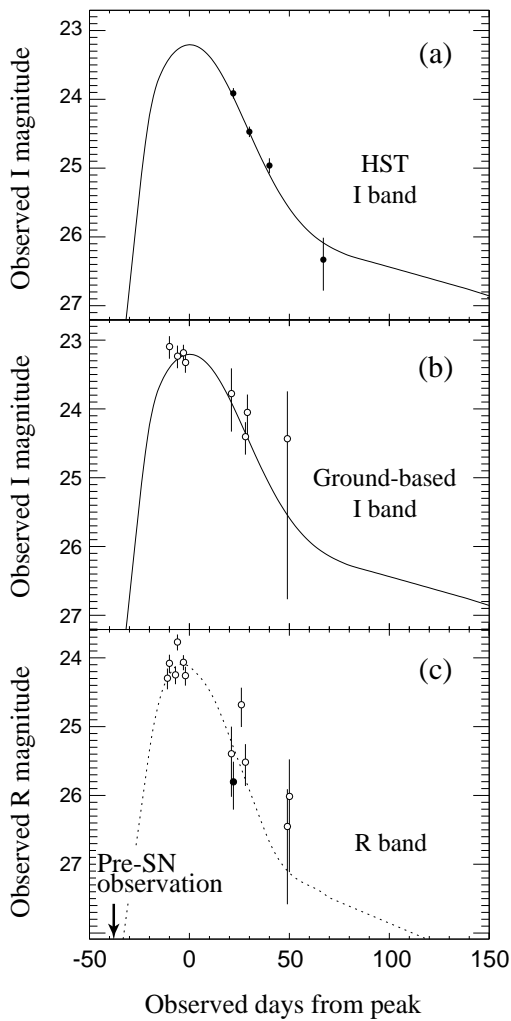,height=5.3in}}
\caption[lvsm] { \small Photometry points for SN 1997ap (a) as
observed by the HST in the F814W filter; (b) as observed with
ground-based telescopes in the Harris $I$ filter; and (c) as observed
with the ground-based telescopes in the Harris $R$ filter (open
circles) and the HST in the F675W filter (filled circle); with all
magnitudes corrected to the Cousins $I$ or $R$ systems$^{13}$.  The
solid line shown in both (a) and (b) is the simultaneous best fit of
the ground- and space-based data to the $K$-corrected, ($1+z$)
time-dilated Leibundgut $B$-band SN~Ia template light
curve$^{22}$\nocite{le:sup}, and the dotted line in (c) is the best
fit to a $K$-corrected, time-dilated $U$ band SN~Ia template light
curve. 
The ground-based data was reduced and calibrated following the
techniques of ref 5, but with no host-galaxy light subtraction necessary.
The HST data was calibrated and corrected for charge transfer
inefficiency following the prescriptions of refs. 23 and 24. K-corrections
were calculated as in ref 25\nocite{akim:kcorr}, modified for the HST
filter system. Correlated zeropoint errors are accounted for in the
simultaneous fit of the lightcurve.  The errors in the calibration,
charge transfer inefficiency correction and K-corrections for the HST
data are much smaller ($\sim$4\% total) than the contributions from
the photon noise. No corrections were applied to the HST data for a
possible $\sim$4\% error in the zeropoints (P.~Stetson, private
communication) or for non-linearities in the WFPC2 response$^{26}$,
which might bring the faintest of the HST points into tighter
correspondence with the best fit lightcurve in (a) and (c). Note that
the individual fits to the data in (a) and (b) agree within their
error bars, providing a first-order cross check of the HST
calibration.}
\label{lightcurve}
\end{figure}

\begin{figure}[tbp]
\centerline{\psfig{figure=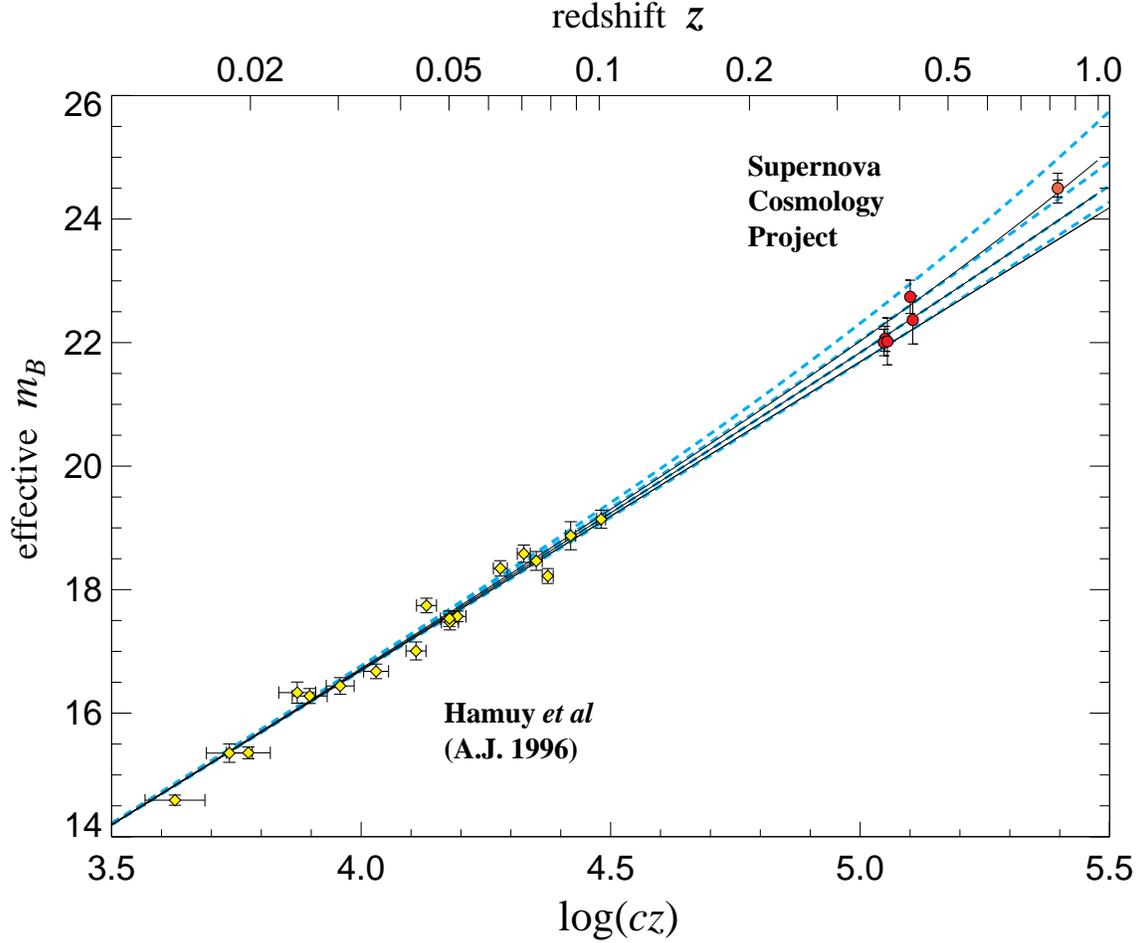,height=5in}}
\caption[lvsm]{SN 1997ap at $z=0.83$ plotted on the Hubble diagram
from ref 5 \nocite{7sne} with the five of the first seven
high-redshift supernovae that could be width-luminosity corrected and
the 18 of the lower-redshift supernovae from the Cal\'an/Tololo
Supernova Survey that were observed earlier then 5 days after maximum
light.  Magnitudes have been $K$-corrected and corrected for the
width-luminosity relation.  The inner error bar on the SN 1997ap point
corresponds to the photometry error alone while the outer error bar
includes the intrinsic dispersion of SNe~Ia after stretch
correction. The solid curves are theoretical $m_B$ for ($\Omega_{\rm
M}$, $\Omega_\Lambda$) = (0, 0) on top, (1, 0) in middle, and (2, 0)
on bottom.  The dotted curves are for the flat universe case, with
($\Omega_{\rm M}$, $\Omega_\Lambda$) = (0, 1) on top, (0.5, 0.5), (1,
0), and (1.5,~$-$0.5) on bottom.}
\label{hubble}
\end{figure}

\begin{figure}[tbp]
\centerline{\psfig{figure=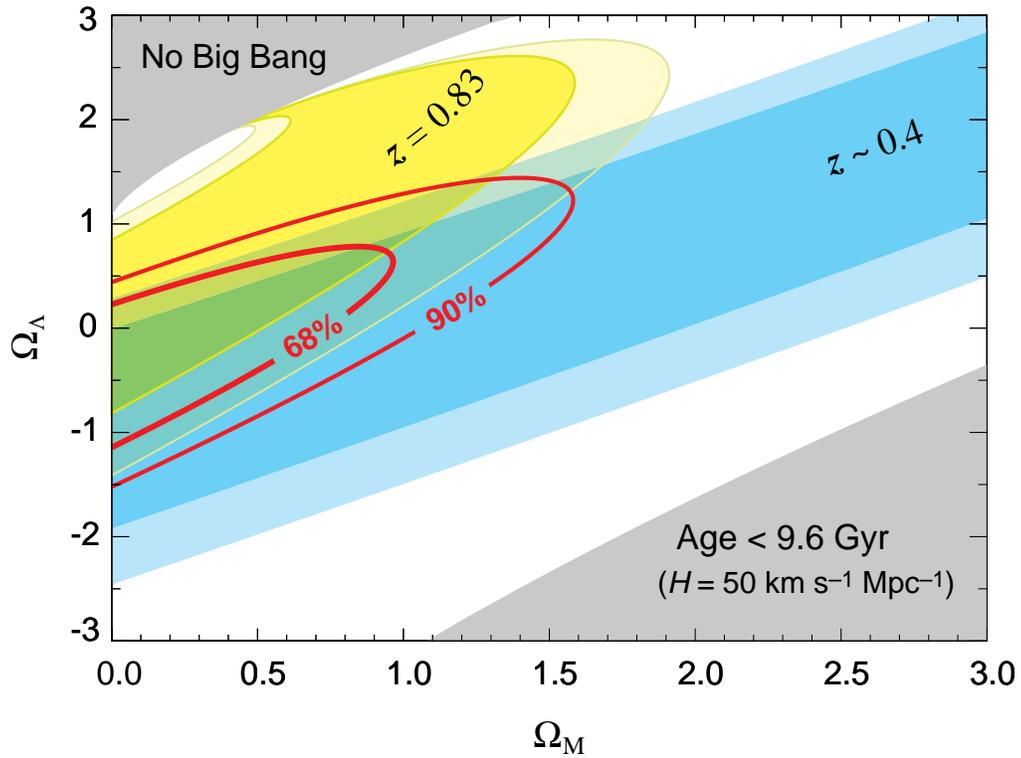,height=5in}}
\caption[lvsm]{[{\it Color Version}] Contour plot of the 68\%
(1$\sigma$) and 90\%, confidence regions in the $\Omega_{\Lambda}$
versus $\Omega_{\rm M}$ plane, for (blue shading) the five supernovae
at $z \sim 0.4$ (see ref 5); (yellow shading) SN 1997ap at $z=0.83$,
and (red contours) all of these supernovae taken together.  The two
labeled corners of the plot are ruled out because they imply: (upper
left corner) a ``bouncing'' universe with no big bang$^{27}$, or
(lower right corner) a universe younger than the oldest heavy
elements, $t_0 < 9.6$ Gyr$^{28}$, for any value of $H_0\ge 50$ km
s$^{-1}$ Mpc$^{-1}$.  }
\label{contour}
\end{figure}

\setcounter{figure}{3}

\begin{figure}[tbp]
\centerline{\psfig{figure=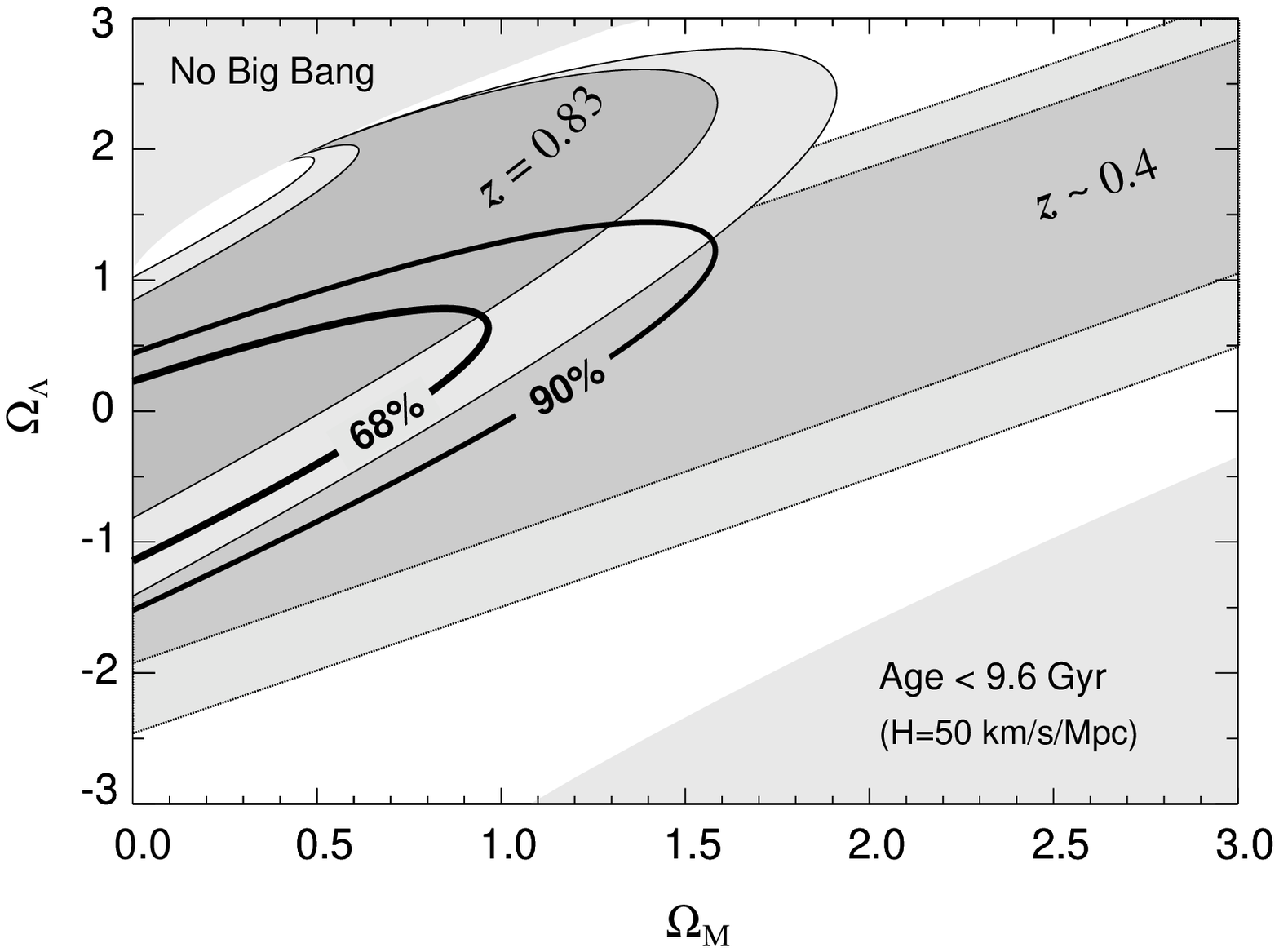,height=5in}}
\caption[lvsm]{ [{\it Black and White Version}] Contour plot of the
68\% (1$\sigma$) and 90\%, confidence regions in the
$\Omega_{\Lambda}$ versus $\Omega_{\rm M}$ plane, for the five
supernovae at $z \sim 0.4$ (see ref 5), SN 1997ap at $z=0.83$, and
(shown as dark, unfilled contours) all of these supernovae taken together.  The
two labeled corners of the plot are ruled out because they imply:
(upper left corner) a ``bouncing'' universe with no big bang$^{27}$,
or (lower right corner) a universe younger than the oldest heavy
elements, $t_0 < 9.6$ Gyr$^{28}$, for any value of $H_0\ge 50$ km
s$^{-1}$ Mpc$^{-1}$.  }
\end{figure}
 

\begin{thebibliography}{10}
\bibitem{ph:delta} Phillips, M.~M. The absolute magnitudes of Type Ia
supernovae. 
\newblock {\em Astrophys.~J.}{ \bf 413}, L105-L108 (1993).

\bibitem{ha:hubble} Hamuy, M., et~al. The Absolute Luminosities of the
 Calan/Tololo Type Ia Supernovae. 
\newblock {\em Astron.~J.}{ \bf 112}, 2391-2397 (1996).

\bibitem{re:lcs} Riess, A.~G., Press, W.~H., and Kirshner, R.~P. Using
Type Ia supernova light curve shapes to measure the Hubble constant. 
\newblock {\em Astrophys.~J.}{ \bf 438}, L17-L20 (1995).

\bibitem{pe:aigua} Perlmutter, S. et~al. Scheduled Discoveries of 7+
High-Redshift Supernovae: First Cosmology Results and Bounds on $q_0$.
\newblock in { \it Thermonuclear Supernovae} (eds P. Ruiz-Lapuente et
al.,) 749-763 (Dordrecht: Kluwer, 1997).

\bibitem{7sne} Perlmutter, S., et~al. Measurements of the Cosmological
Parameters $\Omega$ and $\Lambda$ from the First Seven Supernovae at $z \ge
0.35$.  
\newblock {\em Astrophys.~J.}{ \bf 483}, 565-581 (1997).

\bibitem{pe:ninesn96}
Perlmutter, S. et~al.
\newblock  (1997a).
\newblock International Astronomical Union Circular, no. 6621 and
references therein.

\bibitem{sch:97}
Schmidt, B. et~al.
\newblock  (1997).
\newblock International Astronomical Union Circular, no. 6646 and
references therein.

\bibitem{go:lambda} Goobar, A. and Perlmutter, S. Feasibility of
Measuring the Cosmological Constant $\Lambda$ and Mass Density $\Omega$
Using Type Ia Supernovae. 
\newblock {\em Astrophys.~J.}{ \bf 450}, 14-18 (1995).

\bibitem{sntime} Riess, A.~G., et~al. Time dilation from spectral
feature age measurements of Type Ia supernovae.  
\newblock {\em Astron.~J.}{ \bf 114}, 722-729 (1997).

\bibitem{nugseq95} Nugent, P. et~al. Evidence for a Spectroscopic
Sequence among Type Ia Supernovae. 
\newblock {\em Astrophys.~J.}{ \bf 455}, L147-L150 (1993).

\bibitem{gerson:96} Goldhaber, G. et~al. Observation of cosmological
time dilation using Type Ia supernovae as clocks.
\newblock in { \it Thermonuclear Supernovae} (eds P. Ruiz-Lapuente et
al.,) 777-784 (Dordrecht: Kluwer, 1997).

\bibitem{leib:96} Leibundgut, B. et~al. Time dilation in the light
curve of the distant Type Ia supernova SN 1995K.   
\newblock {\em Astrophys.~J.}{ \bf 466}, L21-L44 (1996).

\bibitem{be:ubvri} Bessell, M.~S. UBVRI passbands.
\newblock {\em Pub. Astr. Soc. Pacific}{ \bf 102}, 1181-1199 (1990).

\bibitem{bran:aigua} Branch, D., Nugent, P. and Fisher, A. Type Ia
supernovae as extragalactic distance indicators.
\newblock in { \it Thermonuclear Supernovae} (eds P. Ruiz-Lapuente et
al.,) 715-734 (Dordrecht: Kluwer, 1997).

\bibitem{b&h} Burstein, D. and Heiles, C. Reddenings derived from H I
and galaxy counts - Accuracy and maps 
\newblock {\em Astron.~J.}{ \bf 87}, 1165-1189 (1982).

\bibitem{pe:fourteensn97}
Perlmutter, S. et~al.
\newblock  (1997c).
\newblock International Astronomical Union Circular, no. 6540.

\bibitem{le:sn1990n} Leibundgut, B., et~al. Premaximum observations of
the Type Ia SN 1990N. 
\newblock {\em Astrophys.~J.}{ \bf 371}, L23-L26 (1991).

\bibitem{we:89b}Wells, L.~A. et~al. The Type Ia supernova 1989B in NGC
3627 (M66). 
\newblock {\em Astron.~J.}{ \bf 108}, 2233-2250 (1994).

\bibitem{br:81b} Branch, D. et~al.The Type I supernova 1981b in NGC
4536: the first 100 days.
\newblock {\em Astrophys.~J.}{ \bf 270}, 123-139 (1983).

\bibitem{pa:94d} Patat, F. et~al. The Type Ia supernova 1994D in NGC
4526: the early phases. 
\newblock {\em Mon. Not. R. Astron. Soc.}{ \bf 278}, 111-124 (1996).

\bibitem{iue:sn}Cappellaro, E., Turatto M., and Fernley J., 
\newblock in {\it IUE - ULDA Access Guide No. 6: Supernovae} (eds
Cappellaro, E., Turatto M., and Fernley J.)
(ESA, Noordwijk, The Netherlands, 1995).

\bibitem{le:sup} Leibundgut, B., Tammann, G., Cadonau, R., and
Cerrito, D. Supernova studies. VII. An atlas of light curves of
supernovae type I. 
\newblock {\em Astro. Astrophys. Suppl. Ser.}{ \bf 89}, 537-579 (1991).

\bibitem{holt:wfpc2} Holtzman, J. et~al. The photometric performance
and calibration of WFPC2. 
\newblock {\em Pub. Astr. Soc. Pacific}{ \bf 107}, 1065-1093 (1995).

\bibitem{whit:wfpc2} Whitmore, B. and Heyer, I. New Results on Charge
Transfer Efficiency and Constraints on Flat-Field Accuracy. 
\newblock {Instrument Science Report WFPC2} {\bf 97-08}, (1997).

\bibitem{akim:kcorr} Kim, A., Goobar, A., and Perlmutter, S. 
 A generalized K-corrections For Type Ia supernovae: comparing R-band
 photometry beyond z=0.2 with B, V, and R-band nearby photometry.
\newblock {\em Pub. Astr. Soc. Pacific}{ \bf 108}, 190-201 (1996).

\bibitem{stee:wfpc2} Stiavelli, M. and Mutchler, M. WFPC2 Electronics 
Verification.  
\newblock {Instrument Science Report WFPC2} {\bf 97-07}, (1997).

\bibitem{carrol:92} Carrol, S., Press, W., and Turner, E. The
cosmological constant.
\newblock Ann. Rev. Astron. Astrophys., {\bf 30}, 499-542 (1992). 

\bibitem{schramm:90} Schramm, D.
\newblock in { \it Astrophysical Ages and Dating Methods} (eds
E.~Vangioni-Flam et~al.,) Gif sur Yvette: Editions Fronti\`eres, 1990)

\end{thebibliography}
\end{document}